\documentclass[aps,prl,twocolumn,showpacs,superscriptaddress,groupedaddress]{revtex4}  
\usepackage{graphicx}  
\usepackage{dcolumn}   
\usepackage{bm}        
\usepackage{amssymb}   
\usepackage{amsmath}   
\usepackage{slashed}
\begin{document}



\title{Extra dimension of space-time exposed by anomalies at low energy}
\author{Nguyen Ai Viet} \affiliation{ITI, Vietnam National University, Hanoi, Vietnam} \affiliation{Physics Department, College of Natural Sciences, Vietnam National University, Hanoi, Vietnam}

\date{\today}

\begin{abstract}
    
Recent experimental observations are shown to be quantitatively consistent with an extended concept of space-time having a discrete extra dimension of two points at the distance of 11.8 fm together with a nontrivial metric structure. In such a space-time, fermions appear in pair with their Kaluza-Klein siblings. The usual electromagnetic field is accompanied with a new vector boson $X17$, which receives a mass of $17~MeV$ from another Kaluza-Klein partner, a scalar boson $H$ of a mass in the range of $0.5-793~keV$ via an abelian Higgs mechanism. At a low energy scale, where nucleons can be treated as structureless in a good approximation, the natural particle model involving nucleons, electron, neutrino and their Kaluza-Klein partners coupled to the electromagnetic field and the massive vector boson $X7$ can lead to new phenomenological consequences, which are verifiable at the currently accessible energy. 
\end{abstract}

\maketitle

{\it Anomalies at nuclear energy scale}--
In 2016, Krasznahorkay et al. \cite{ATOMKI1} have observed an anomalous internal pair conversion (IPC) in the excited beryllium nucleus transition to its ground state $^8Be^*(18.15~MeV) \rightarrow ^8Be(g.s) $ when bombarding a $^7Li$ target with a low energy proton beam. The anomaly can be explained by assuming the existence of a vector boson of mass $17~MeV$ ($X17$), which is released in place of photon from the excited state of energy $18.15~MeV$. In average, there are consistently $5.8$ events related to $X17$ in a total of 1 million ones, mostly dominated by photon. The measured data can be interpreted as the following relation
\begin{equation} \label{ATOMKI}
{\Gamma (^{8}Be^* ~\rightarrow~ ^{8}Be+X17) \over 
	\Gamma (^8Be^* \rightarrow~ ^{8}Be + \gamma)} B = 5.8 \times 10^{-6},
\end{equation}
where $B$ is the branching ratio of the $X17 \rightarrow e^+ + e^-$ decay channel. 

Naturally, one implies that the interaction of $X17$ with the ordinary matter is about thousand times weaker than that of photon. As a consequence, the authors have interpreted $X17$ as the dark photon. The experiment has been repeated with different settings and refined methods and have always given the high confidence of $6.8 \sigma$ \cite{ATOMKI2}. Recently, new observations of $X17$ have also been shown  in the transition of the $^4He (20.6~MeV)$ excitation to the ground state \cite{ATOMKI3}. Feng et al. \cite{Feng1, Feng2} have carried out a careful theoretical analysis on the ATOMKI's first experiment and concluded that it is about the new fifth interaction, which is "protophobic", meaning that $X17$ interacts with proton much more feebly than with neutron. Other research groups have also postulated different semi-empirical models to explain this anomaly. In any cases, the vector boson $X17$ with a mass in the $MeV$ range can be a possible portal to a new sector, which is currently in the intensive search of different research groups \cite{Ban}, paving the way to new particles not known in the Standard Model. However exciting the perspective of $X17$ is, the question remains: why it is there and in what relation to photon? If such a significant fifth interaction exists, its existence must be based on some fundamental principles as all other interactions do in Einstein's General Relativity and the Standard Model. 
   
On the other hand, the neutron life time puzzle can also shed light on the existence of $X17$. It has been known for many years that the life times of neutron measured by different methods are in discrepancy beyond the experimental errors. Recently, Fornal and Grinstein \cite{FoGr} have proposed a new resolution to this issue by assuming that the neutron has additional decay channels into the dark sector beyond the known $\beta$ one. The authors have suggested three different channels: i) $n \rightarrow n_X + \gamma$, ii) $n \rightarrow n_X + e^+ + e^-$ and iii) $n \rightarrow n_X + \phi$, where $n_X$ is the dark neutron and $\phi$ is a dark scalar particle. The recent experiment tests have practically excluded the first two channels \cite{Tang, Sun}. Without additional detailed information about $n_X$ and $\phi$ the channels iii) can neither be confirmed nor falsified. One can wonder if $X17$ can mediate a new interaction and trigger the new decay channels in addition to the channel $ n \rightarrow n_X+ \nu + \bar \nu$ as proposed by Ivanov et al. \cite{Ivanov2019}.
 
Our intension is to probe into a new physical space-time, which is an extension of the traditional  ${\cal M}^4$ with an additional discrete extra dimension consisting of two points separated from each other by a distance $d \sim 11.8~fm$. This size is equivalent to the inverse of the mass $m = 17~MeV$. In such a space-time, the physical fields appear together with their Kaluza-Klein (KK-) partners. The massive vector field $X_\mu(x)$ is a KK-partner of the massless photon field $A_\mu(x)$. Neutron and the KK-neutron $n_X$ are also siblings in a pair. We are able to show that this picture is consistent quantitatively with the observational data related to the ATOMKI's experiment and neutron decay. Moreover, these experimental data implies strong constraints on the parameters of the  particle model.
 
{\it Extra dimensions and discrete extra dimensions}.-- 
The large extra dimension (LED) proposed by Arkani-Hamed, Dimopoulos and Dvali \cite{ADD} postulates a large extra dimension of size $1~mm$ to bring the value of the Planck mass to the $TeV$ energy scale. The size of LED can also be chosen around 1 fermi, if several LEDs are considered. The universal extra dimension (UED) proposed by Appelquist, Cheng and Dobrescu \cite{UED}, postulates a size of $10^{-18} m$  much larger than the traditional Planck scale but smaller than LED, to bring the Planck scale down to $1000 TeV$. The Randall-Sundrum model (RS1) \cite{RS1} also postulates an extra dimension with two specific branes of $TeV$ and Planck energy scales with a warping factor to solve the hierarchy problem. Both LED and RS1 require an assumption that the physical fields are localized on some membranes. 

The discrete extra dimension (DED) is an alternative proposal, which is originated from Connes-Lott's model \cite{CoLo} of two sheeted space-time where the right- and left-handed chiral fermion particles of the Standard Model exist. The discretized Kaluza-Klein theory (DKKT) with generic DEDs having just two points has been developed by Viet and Wali since 1994 \cite{LVW, VW1995a, VW1995b, VW2000,VW2003, VietDu, VDHW} in the extended space-time based on the general mathematical foundation of noncommutative geometry (NCG) originally proposed by Connes \cite{Connes}. DKKT utilizes a more intuitive formulation, which is in parallelism with the traditional Einstein-Cartan and Kaluza-Klein theories \cite{Kalu, Klein}. The main advantage of DKKT is that it does not need a cumbersome treatment of the infinite towers of massive modes, which leads to both experimental and theoretical inconsistencies. The concept of discrete extra dimensions has also been investigated by other authors \cite{Alis, ArSch2004, DeMo2005}. In particular, it has been suggested that the Kaluza-Klein partners of the ordinary fields in DKKT  can be candidates of a new matter type \cite{QuyNhon2015}. This discrete dimension is necessary to unify all the interactions and the Higgs field as components of the extended gravity. The discrete extra dimension has also been shown to be relevant beyond the high-energy physics, in particular in the bilayer Quantum Hall systems \cite{JVWQHE}. Since the KK-siblings in DKKT is always a subset of the spectrum in the space-time with a continuous extra dimension, our result can be considered as the first test of the general idea of  extra dimensions at low energy.

{\it The extended space-time}.-- Our particular space-time is an extension of the usual one ${\cal M}^4$ with a discrete extra dimension having the following specific line element and metric
\begin{eqnarray}\label{Line}
ds^2 &=& G_{MN} dx^M dx^N = \eta_{\mu \nu} dx^\mu dx^\nu 
+ \lambda^4 dx^5 dx^5, \nonumber \\
G_{\mu \nu}&=& diag(-1,1,1,1), \nonumber \\ G_{\mu 5}&=& G_{5\mu}=0, ~~ G_{55}= \lambda^4,
\end{eqnarray} 
where $M,N = \mu, 5$.

In DKKT \cite{VW1995b}, the most general extended metric contains Kaluza-Klein pairs of metrics and vector fields defined on two space-time copies together with a Brans-Dicke scalar, which measures the varied distance between the two sheets at each given point. The line element defined in Eq.(\ref{Line}) is a specific metric structure of DKKT with two flat metrics, two vanishing vector fields and a Brans-Dicke scalar being frozen to a constant value $\lambda^2$ to be determined by experimental data.

The traditional Hilbert-Einstein action integral has an invariant volume element $dx^4 \sqrt{-det g}$. DKKT with the metric in Eq.(\ref{Line}) has the extended volume element $dx^4 \sqrt{-det g}\lambda^2 = dx^4 \lambda^2$. That is to say, the constant Brans-Dicke scalar add a factor $\lambda^2$ to the Lagrangian of our model. We will see that the metric parameter $\lambda$ will play a crucial role in our model to give quantitatively reasonable results. 

The extended vielbein $E^A_M(x)$ and its inverse $E^M_A(x)$ normalizes the non-normal basis $DX^M$ to the locally flat one as follows
\begin{subequations}
\begin{alignat}{3}
&E^A = E^A_M DX^M, ~~ DX^M = E^M_A E^A&  \\
&G^{AB} = E^A_M G^{MN} E^B_N = diag(-1,1,1,1,1)& \\
& G^{MN} G_{NL} = \delta^M_L
\end{alignat}
\end{subequations}
where $A,B= a, {\dot 5}$ are the index of the locally orthonormal frame.

For the non-normal metric in Eq.(\ref{Line}) we have the only non-trivial vielbein components
\begin{equation}
E^5_{\dot 5} = \lambda^{-2},~~E_5^{\dot 5} = \lambda^2.
\end{equation}


{\it Extended fermion and vector fields}.--
It is convenient to represent a given fermion Kaluza-Klein pair as a 2-column spinor
\begin{equation}\label{spinor}
\Psi =
\begin{bmatrix}
& \psi_1 & \\
& \psi_2 &
\end{bmatrix}.
\end{equation}
 
The usual Dirac operator $\slashed {\partial}= \gamma^\mu \partial_\mu$ now is extended to the following matrix
\begin{equation}\label{Doperator}
\slashed{D} = \slashed{\partial}. {\bf 1} + \Theta =
\begin{bmatrix}
& \slashed{\partial} & -i m\theta  \\
& i m \theta & \slashed{\partial}
\end{bmatrix},
\end{equation}
where $\theta$ ($\theta^2=1$) is a Clifford element, which is an analogue of Dirac matrices in the fifth dimension, while $m$ is a mass parameter.

Additionally, we also require the action of $\theta$ on the spinors as follows
\begin{equation} \label{darkcharge}
\theta \psi_1 = \psi_1, ~~ \theta \psi_2 = -\psi_2
\end{equation}

We have two sets of generalized Dirac matrices $\Gamma^A = \Gamma^M E^A_M$ and $\Gamma^M = \Gamma^A E^M_A$ satisfying the trace relations
\begin{equation}
Tr(\Gamma^M \Gamma^N) = G^{MN},~~ 
Tr(\Gamma^A \Gamma^B) = G^{AB}
\end{equation}

In our specific metric we have
\begin{eqnarray}
\Gamma^\mu &=& \delta^\mu_a \Gamma^a = \begin{bmatrix}
\gamma^\mu & 0 \cr
0 & \gamma^\mu 
\end{bmatrix}, \nonumber \\
\Gamma^5 &=& E^5_{\dot 5} \Gamma^{\dot 5} = \begin{bmatrix}
0 & i \theta/\lambda^2 \cr
-i \theta /\lambda^2 & 0
\end{bmatrix}.
\end{eqnarray}.

The Kaluza-Klein partners of photon are obtained by extending the usual vector field's 1-form $\slashed {b} = \gamma^\mu b_\mu(x)$  into the hermitian  $2 \times 2$ matrix 1-form operator as follows    
\begin{eqnarray}
\slashed{B} &=& \Gamma^\mu B_\mu(x) + \Gamma^5 \phi(x) =
\begin{bmatrix}
\slashed{b}_1  &  i \theta \varphi(x)/\lambda^2 \\
- i \theta \varphi(x)/\lambda^2 & \slashed{b}_2
\end{bmatrix} \nonumber \\
&=& \begin{bmatrix}
g \slashed{A} Q + g' \slashed X Q_X &  i \theta \varphi(x)/\lambda^2 \\
- i \theta \varphi(x)/\lambda^2 & g\slashed{A} Q
\end{bmatrix},
\end{eqnarray}
where $A_\mu(x)$ is the usual electromagnetic field. $X_\mu(x)$ is the vector Kaluza-Klein partner of $A_\mu(x)$. $Q$ and $Q_X$ are the charge and X-charge operators. These charges are the same for the fermion states $\psi_1$ and $\psi_2$. $g$ and $g'$ are coupling constants related to photon and $X17$ respectively. Thus, the extended vector field consists not only of a pair of vector fields $A_{\mu}(x)$ and $X_{\mu}(x)$, but also a scalar field $\varphi(x)$, which actually is the fifth component of the extended electromagnetic field. 

{\it Abelian Higgs mechanism}-- The generalized field strength 2-form is defined as follows
\begin{equation}
{\bf B} = \slashed {D} \wedge \slashed{B} + \slashed{B} \wedge \slashed{B} = \Gamma^M \wedge \Gamma^N B_{MN},
\end{equation}
where the extended wedge product is defined as
\begin{eqnarray}\label{Wedge2}
\Gamma^\mu {\wedge} \Gamma^\nu & =& {1 \over 2} [\Gamma^\nu, \Gamma^\mu],~~ \Gamma^5 {\wedge} \Gamma^5  \not=  0 \nonumber \\
\Gamma^\mu \wedge \Gamma^5 & = & - \Gamma^5 {\wedge} \Gamma^\mu = \Gamma^\mu \Gamma^5.
\end{eqnarray}

The components of the field strength tensor $B_{MN}$ are calculated as follows
\begin{subequations}\label{Field}
	\begin{align}
	&B_{\mu \nu} =  {1 \over 2} (\partial_\mu B_\nu(x)- \partial_\nu B_\mu(x)) \hskip 1.8cm&\\
	&B_{\mu 5} = {1 \over 2} (\partial_\mu + g' X_\mu(x) Q_X 
	\begin{bmatrix}
	1 & 0 \\
	0  & -1
	\end{bmatrix}	
	) (\varphi(x) + m) &\\
	&B_{55} = 2m \varphi(x) + \varphi^2(x). ~\hskip 2.8cm& 
	\end{align}
\end{subequations}
We define the physical scalar field $h(x)$, the field strength tensors of the vector fields $A_\mu(x)$ and $X_\mu(x)$ as follows
\begin{subequations}
\begin{alignat}{3}
&\varphi(x) = f_\kappa \lambda h(x) - m& \\
&F_{\mu \nu} = \partial_\mu A_\nu(x) - \partial_\nu A_\mu(x), & \\
&X_{\mu \nu} = \partial_\mu X_\nu(x) - \partial_\nu X_\mu(x).&
\end{alignat}
\end{subequations}

The action for the gauge sector now is
\begin{widetext}
	\begin{eqnarray}
	{\mathcal S}_{g} &=& - { 1 \over 2f^2_\kappa N_f}\int d^4x  Tr (B^{\mu \nu} B_{\mu \nu} \lambda^2 + 2 B_{\mu 5} B_{\nu 5} \lambda^{-2} 
	+  B^2_{55} \lambda^{-6}), \nonumber \\
 &=& \int d^4x ( - {1\over 4} F^{\mu \nu} F_{\mu \nu} - {1 \over 4} X^{\mu \nu} X_{\mu \nu}+ {1 \over 2 } \partial^\mu h(x) \partial_\mu h(x) + f^2_\kappa \lambda^2 X^2_\mu(x) h^2(x) - f^2_\kappa \lambda^{-2} (h^2(x) - {m^2 \over f^2_\kappa \lambda^2})^2). \label{Gauge}
	\end{eqnarray}
\end{widetext}

Motivated by by the proposal of Mohapatra and Marshak \cite{MoMa}, we  identify $Q_X$ as the $B-L$ number operator and obtain $\sum Q^2_X = N_f$, where $N_f$ is the number of fermions in the given model. Therefore, we must choose the parameters $f_\kappa$ and $g'$ to satisfy the following relations to have the correct factors for the kinetic terms 
\begin{equation}
2g^2 \sum Q^2 = g'^2 N_f,~ f_\kappa = g' \lambda,
\end{equation}
where the sum is taken over all the fermions in the model. 

Due to its quartic potential term, the scalar field $h(x)$ has a vacuum expectation value (VEV) $v= m/f_\kappa \lambda$. 
The vector field $X_\mu(x)$  receives a mass from the VEV of the Higgs field $h(x)$. The mass of the scalar Higgs field $h(x)$ is $ \sqrt{2} m/\lambda^2$.

{\it Fermion mixing and mass splitting}--
The usual action of the free massive spinor interacting with a vector field $\int dx^4 \bar \psi (i (\slashed{\partial} + \slashed{b}) + \mu_f) \psi$ now is extended to the following action
\begin{widetext}
\begin{subequations} \label{goodlag}
\begin{alignat}{3}
	{\mathcal S}_f &=& \int dx^4 \lambda^2 Tr( i {\bar \Psi} (\slashed{D} + \slashed{B}) \Psi +  {\bar \Psi} 
	\begin{bmatrix}
	\mu_f & 0 \\
	0 & \mu'_f
	\end{bmatrix}
	\Psi) = {\mathcal S}_{f-g}+{\mathcal S}_{f-h}, \hskip 1.1cm  \\
{\mathcal S}_{f-g}	&=& ~~\int dx^4 i Tr(  {\bar \Psi}' (\slashed{D} + \slashed{B}) \Psi' = \int dx^4 (i {\bar \psi}'_1 (\slashed{\partial}+\slashed{b}_1 )\psi'_1 + i {\bar \psi}'_2( \slashed{\partial}+ \slashed{b}_2 ) \psi'_2 )  \\
{\mathcal S}_{f-h} &=& \int d^4x Tr(  {\bar \Psi}' {\cal M} \Psi') 
= \int d^4x  Tr({\bar \Psi}' \begin{bmatrix}
	\mu_f &  f_\kappa h(x)/\lambda \\
	f_\kappa h(x)/\lambda  & \mu'_f 
	\end{bmatrix}
	\Psi'). \hskip 0.9cm
\end{alignat}
\end{subequations}
\end{widetext}
The constant $\lambda^2$ has been absorbed in the spinor wave function by redefinition $\Psi' = \Psi /\lambda $. $\mu_f$ and $\mu'_f$ are masses of the fermionic pair without interaction with the gauge sector. From now on, the prime on the redefined wave functions will be omitted.

When the scalar field $h(x)$ is frozen to the VEV $v$, we will have a non-diagonal mass matrix ${\mathcal M}$. To obtain the mass eigenstates, we have to diagonalize ${\mathcal M}$ by a unitary transformation.

For convenience, we replace the mass variable $\mu'_f$ by introducing a new variables $\theta$  defined by the following relation
\begin{equation}
\cot 2\theta = {(\mu'_f - \mu_f) \over 2m}.
\end{equation}   

Now the mass matrix can be diagonalized by the $2 \times 2$ unitary transformation $U$ with the mixing angle $\theta$ as follows 
\begin{widetext}
\begin{subequations}
	\begin{alignat}{3}
	\Psi_U &=&~\begin{bmatrix}
	& \psi & \\
	& \psi_X &
	\end{bmatrix} = U \Psi = 
	\begin{bmatrix}
	~~\cos \theta & \sin \theta \cr
	- \sin \theta & \cos \theta 
	\end{bmatrix} \Psi, \hskip 3.1cm \\
	{\cal M}' &=&~~ U  {\cal M} U^\dagger = 
	\begin{bmatrix}
	 \mu_f-  h(x)\tan f_\kappa\theta /\lambda  & 0 \cr
	0 & \mu_f + h(x) f_\kappa\cot \theta /\lambda 
	\end{bmatrix}.
	\end{alignat}
\end{subequations}

The action term ${\mathcal S}_{f-h}$ now is  expressed in terms of the mass eigenstates $\psi$ and $\psi_X$ as follows

\begin{equation} \label{Fmass}
{\cal S}_{f-h} = \int d^4x {\bar \psi} (\mu_f - h(x) \tan \theta f_\kappa/\lambda ) \psi  
+  {\bar \psi}_X (\mu_f + h(x) \cot \theta f_\kappa /\lambda ) \psi_X 
\end{equation} 
\end{widetext}

The Kaluza-Klein siblings of a given fermion $\psi$ and $\psi_X$  have the same "bare" mass $\mu_f$ and receive different additional mass correction from the scalar $h(x)$ due to the abelian Higgs mechanism. The mass splitting between these fermions is
\begin{equation}
\delta m_\psi = m_{\psi_X} - m_\psi = \frac{2m}{\sin 2 \theta_\psi \lambda^2 }. \label{msplit}
\end{equation}

The action ${\mathcal S}_{g-f}$ of the mass eigenstates $\psi$ and $\psi_X$ coupled to the gauge sector is as follows
\begin{widetext}
\begin{eqnarray}
{\mathcal S}_{f-g}&=& i\int d^4x \bar Tr \Psi_U ( \slashed{\partial} + U 
\begin{bmatrix}
g\slashed {A} Q + g'  \slashed X Q_X & 0 \\
0 & g\slashed {A} Q 
\end{bmatrix} U^\dagger)
\Psi \nonumber \\
&=& i\int d^4x \bar \Psi_U ( \slashed{\partial} + 
\begin{bmatrix}
g\slashed {A}Q + g' \cos^2 \theta \slashed{X} Q_X & -1/2 g'\sin 2\theta  \slashed{X} Q_X   \\
- 1/2 g' \sin 2\theta \slashed{X} Q_X & g\slashed {A} Q + g' \sin^2  \theta\slashed{X} Q_X
\end{bmatrix}
 \Psi_U \nonumber\\
 &=& i\int d^4x (\bar \psi (g\slashed {A} Q + g' \cos^2 \theta \slashed{X} Q_X ) \psi - \frac{g'}{2} \bar \psi_X  \sin 2\theta \slashed{X} Q_X \psi - \frac{g'}{2} \bar \psi \sin 2\theta \slashed{X} Q_X \psi_X \nonumber \\
 && + \bar \psi_X (g\slashed {A} Q + g' \sin^2 \theta \slashed{X} Q_X ) \psi_X ).  \label{Fgauge}
\end{eqnarray}
\end{widetext}

{\it The extended particle model of nuclear physics} --  
The minimal fermionic sector in our extended electromagnetic model consists of neutron $n$, proton $p$, electron $e^-$, neutrino $\nu_e$ and their KK-partners $n_X, p_X, e^-_X, \nu_X$.  The mass splittings within each particle types are determined by the mixing angles $\theta_n, \theta_p, \theta_e$ and $\theta_\nu$. Although at higher energy like the electroweak scale, the particle model must be constructed out of quark-leptons, at the energy range under $100~MeV$, the structureless nucleons are good approximations.

Now we can finalize the model based on the charges of the particles. In our model $ N_f = 4, \sum Q^2 = 2$ we have
\begin{equation}
g'= g 
\end{equation}

Assuming the Higgs field is frozen to its VEV, we obtain the following particle model

\begin{widetext}
\begin{subequations}\label{KKfield}
\begin{alignat}{2}
{\mathcal S} =&~~~ \int d^4x ({\mathcal L}_g + {\mathcal L}_{n} + {\mathcal L}_{p} +  {\mathcal L}_{e} +  {\mathcal L}_\nu), \\
{\mathcal L}_{g} =& -~{1 \over 4} F^{\mu \nu} F_{\mu \nu} - {1 \over 4} X^{\mu \nu} X_{\mu \nu} + m^2 X^2_\mu(x)  \\
{\mathcal L}_n =&~~~  {\bar n}(x) (\slashed{\partial} + g \cos^2 \theta_n \slashed {X} + m_n ) n(x)  
+  {\bar n}_X(x) ( \slashed{\partial}  + g \sin^2 \theta_n \slashed {X} + m_{n_X}) n_X(x) \nonumber \\
& - {g \sin 2\theta_n \over 2} (\bar n_X(x) \slashed{X} n(x) + \bar n(x)  \slashed{X} n_X(x)), \\
{\mathcal L}_p =&~~~  {\bar p(x)} (\slashed{\partial} + g\slashed{A} + g \cos^2 \theta_p \slashed {X} + m_p) p(x)  
- {g \sin 2\theta_p \over 2} (\bar p_X(x) \slashed{X} p(x) + \bar p(x)  \slashed{X} p_X(x)) \nonumber \\
& ~ 
+  {\bar p}_X(x) ( \slashed{\partial} + g\slashed{A} + g\sin^2 \theta_p \slashed {X} + m_{p_X} ) p_X(x) 
,  \\
{\mathcal L}_e =&~~~  {\bar e(x)} (\slashed{\partial} - g\slashed{A} - g \cos^2\theta_e \slashed {X} + m_e ) e(x)  
+ {g \sin 2\theta_e \over 2}( \bar e_X(x) \slashed{X} e(x) + \bar e(x)  \slashed{X} e_X(x)) 
\nonumber \\
& ~ 
+  {\bar e}_X(x) ( \slashed{\partial} - g\slashed{A} - g \sin^2\theta_e \slashed {X} + m_{e_X}) e_X(x) 
, \\
{\mathcal L}_\nu =&~~~  {\bar \nu}(x) (\slashed{\partial} - g \cos^ 2\theta_\nu \slashed {X} + m_\nu) \nu(x)  
+  {\bar \nu}_X(x) ( \slashed{\partial}  - g \sin^2\theta_\nu \slashed {X} + m_{\nu_X} ) \nu_X(x)  \nonumber \\
&  + {g \sin 2\theta_\nu \over 2}( \bar \nu_X(x) \slashed{X} \nu(x) + \bar \nu(x)  \slashed{X} \nu_X(x)). 
\end{alignat} 
\end{subequations}	
\end{widetext}	

Thus the particle model given in Eqs.(\ref{KKfield}) and the existence of the vector boson $X17$ are rooted in DED characterized by the mass $m=17~MeV$. DED also implies the existence of the KK-siblings of the usual fermions, which opens up new decay channels for $X17$ and neutron. The quantitatively measured decay rates can help to determine the mixing angles, which can fix the unknown mass splittings between the KK-siblings. Therefore, this model is very predictive as we will see later in this article.

{\it Interpretations of the ATOMKI's and neutron decay related observations}-- The ratio between the $\gamma$ and $X17$ related decay channels in Eq.(\ref{ATOMKI}) can be estimated in our particle model in a good approximation as follows
\begin{widetext}
\begin{equation}
{\Gamma (^8Be^* \rightarrow~ ^{~8}Be+X17)
	\over 
	\Gamma (^8Be^* \rightarrow ~^8Be + \gamma)}= (\cos^2\theta_p + \cos^2\theta_n)^2 (1- {m^2 \over E^2(^8Be^*)})^{3/2} \sim 0.043 (\cos^2 \theta_p+ \cos^2\theta_n)^2,
\end{equation}
\end{widetext}
where $E(^8Be^*)=18.15~MeV$ is the energy of the excited state of $^8Be$ nucleus.

So Eq.(\ref{ATOMKI}) leads to the following constraint on our model's parameters
\begin{equation}
(\cos^2 \theta_p+ \cos^2\theta_n)^2 B = 1.35 \times 10^{-4}.
\end{equation} 

In addition to the usual $\beta$-decay, we can have the new neutron decay channels into KK-neutron and a pair of leptons, being mediated by a virtual $X17$ as depicted in FIG I.

\begin{figure}[h]
	\includegraphics[width=\linewidth]{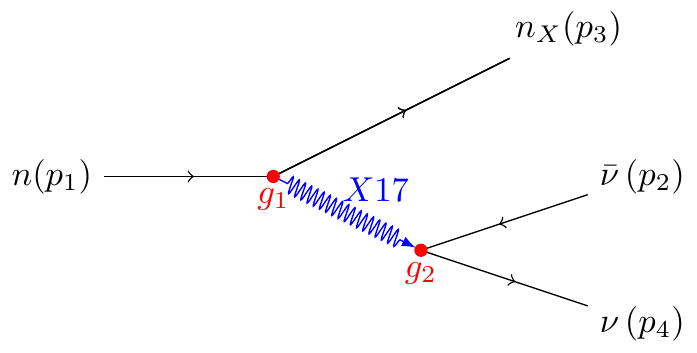}
	\caption{Decay $n \rightarrow n_X + \nu + \bar \nu $ via X17}
	\label{fig:darkneutrondecay}
\end{figure}

The observation of Tang et al \cite{Tang} that the neutron decay is not accompanied with the $\gamma$ radiation is a direct consequence of this model. In order to explain the lack of $e^+-e^-$ pair creation in the neutron decay observed by Sun \cite{Sun}, we can assume
\begin{equation} \label{newconstraint}
\delta m_n = m_n - m_{n_X} <2 m_e= 1.102~MeV,
\end{equation}
which is stronger than the condition suggested previously by Fornal and Grinstein \cite{FoGr} $m_n - m_{n_X} < 1.664 MeV$. The new condition (\ref{newconstraint}) also guarantees the experimental observation that $^9Be$ nucleus is stable against the neutron decay channel $ n \rightarrow n_x + e^+ + e^-$.
 
As a consequence of the condition (\ref{newconstraint}), we obtain the first lower bound for the geometric parameter $\lambda^2$ as follows
\begin{equation}
1 > |\sin2\theta_n | =  \frac {2m}{\delta m_n \lambda^2}  > \frac{30.853}{\lambda^2},~~ \lambda^2 > 30.853.  \label{smallpmixing} 
\end{equation}

We can also reexamine the resolution proposed by Fornal and Grinstein \cite{FoGr} by introducing KK-neutron $n_X$ and the neutrino siblings as final products of the neutron decay instead of the unknown dark particles. The branching ratio of total alternative neutron decay channels into $n_X$ must be around $1\%$ to solve the neutron life time puzzle \cite{FoGr}. That is to say,
\begin{equation} \label{99}
\Gamma_\beta / \Gamma_{n_X} \sim 99,
\end{equation}
where $\Gamma_{n_X}$ is the total decay width of the neutron decay into $n_X$. The $\beta$ decay width $\Gamma_\beta$ is given in the Standard Model \cite{Grif2008} as follows
\begin{eqnarray}
\Gamma_\beta &=& {1.633 m^5_e \over 2 \pi^3} (V_{ud})^2 G^2_F (1 + \frac{3g^2_A}{g^2_V}) \nonumber \\
&=& \frac{9.504 g^4}{ 64 \pi^3 \sin^4 \theta_W} \frac {m^5_e} {m^4_W},  
\end{eqnarray}
$G_F, g_A, g_V$ and $g$ are respectively the Fermi, axial vector, vector weak coupling constants and the electromagnetic one. $\theta_W$ is the Weinberg angle, $m_W \sim 80~ GeV$ is the $W$ boson mass, $V_{ud}$ is the CKW matrix element.

The above constraint on the mass splitting of neutron suppresses all the decay channels related to electron and its KK-sibling. Therefore, the only remaining decay channels are
\begin{subequations} \label{nnxdecay}
	\begin{alignat}{3}
	&n \rightarrow n_X + \nu + \bar \nu, & \label{nudecaya}\\
	&n \rightarrow n_X + \nu_X + \bar \nu, &\\
	&n \rightarrow n_X + \nu + \bar \nu_X, &\\
	&n \rightarrow n_X + \nu_X + \bar \nu_X. &
	\end{alignat}
\end{subequations} 

Depending on the KK-neutrino mass $m_{\nu_X}$ only certain channels are allowed energetically. In a good approximation, the total decay width $\Gamma_{n_X}$ of the $n_X$ related decay channels is given as 
\begin{eqnarray}
\Gamma_{n_X} &\sim&  \frac{g^4 \sin^2 2\theta_n (m_n- m_{n_X})^5 }{60 \pi^3 m^4} C \nonumber \\
&\sim& - \frac{8 g^4 m}{15 \pi^3 \sin^3 2 \theta_n \lambda^{10}} C,
\end{eqnarray}
where $C <1$ and:

  + $C=1$ if $m_{\nu_X} < m_e$, 
  
  + $C= (1 - \sin^4 \theta_\nu)$ if $  m_e < m_{\nu_X} < 2 m_e$,
  
  +  $C=\cos^4 \theta_\nu $ if $m_{\nu_e} > 2 m_e$.\\

The Fornal-Grinstein resolution constraint in Eq.(\ref{99}) now becomes
\begin{equation}
C = - 3.8342 \times 10^{-24} \lambda^{10} \sin^3 2 \theta_n 
\end{equation}

{\it Protophobia of $X17$}-- If we assume that the KK-proton has a mass at least in the $TeV$ range, then  the following condition for proton mixing angles holds
\begin{equation}
\sin 2 \theta_p = \frac{2m}{\lambda^2 \delta m_p} < \frac{3.4 \times 10^{-5}}{\lambda^2} < 1.102 \times 10^{-6}.   \label{largepmass}
\end{equation}

To satisfy Eq.(\ref{largepmass}), we have two alternatives according to the sign of $\cos 2\theta_p$ in the following relation 
\begin{equation}
\cos 2 \theta_p = 2 \cos^2 \theta_p -1 \sim \pm 1.
\end{equation}

Since, the positive sign lead to $\cos^2 \theta_p =1$, meaning that proton interacts equally with $X17$ and photon, which might lead to the observational obstacles, we choose 
\begin{equation}
\cos 2 \theta_p \sim -1,~\cos^2 \theta_p \sim 0,~ \textrm{~when~} \theta_p \sim \pi,
\end{equation}
which means that $X17$ is protophobic as suggested by Feng et al. \cite{Feng1}.

The ATOMKI's constraint in the case of protophobic $X17$ is as follows
\begin{equation} \label{protophobia}
\cos^4 \theta_n B = 1.35 \times 10^{-4}.
\end{equation} 

{\it Small neutrino mass splitting}--
Now we consider the case, where $m_{\nu_X} < m_e$ and $C=1$. The decay widths of all possible decay channels are given by the two-body decay formula \cite{Grif2008} as follows

\begin{widetext}
	\begin{subequations}\label{X17decay}
		\begin{alignat}{2}
		\Gamma_e(X17)=&~\Gamma(X17 \rightarrow e^+ + e^-) = \frac {g^2\cos^4 \theta_e} {32 \pi m}  \sqrt{1- 4 m^2_e/m^2} = 0.998  \frac {g^2\cos^4\theta_e} {32 \pi m}  \label{epair}\\
		\Gamma_\nu(X17,1) =&~\Gamma(X17 \rightarrow \nu + \bar \nu) = \frac {g^2\cos^4\theta_\nu} {32 \pi m}   \label{nupair}\\
		\Gamma_\nu(X17,2)=&~\Gamma(X17 \rightarrow \nu_X + \bar \nu) = \Gamma_\nu(X17, 3)=\Gamma(X17 \rightarrow \nu + \bar \nu_X) \sim \frac{g^2 \sin^2 2\theta_\nu}{32 \pi m}, \label{1nuX} \\
		\Gamma_\nu(X17,4)=&~ \Gamma(X17 \rightarrow \nu_X + \bar \nu_X) =\frac {g^2\sin^4 \theta_\nu} {32 \pi m}, \label{fXpair}   
		\end{alignat}
	\end{subequations}
\end{widetext}	
whose total sum will be independent of the neutrino mixing angle
\begin{equation}
\Gamma_\nu(X17,\nu) =\sum \Gamma_\nu(X17,i)= \frac{g^2}{32 \pi m} 
\end{equation}

Now the branching ratio $B$ for the small neutrino mass splitting is given by the following formula
\begin{equation}
B = \frac{\cos^4 \theta_e}{\cos^4 \theta_e + 1.002},
\end{equation}
which implies from Eq.(\ref{protophobia}) implies that in this scenarios $X17$ cannot be electrophobic since 
\begin{equation}
\cos^2\theta_e > 1.162 \times 10^{-2}.
\end{equation}

So, in order to keep the large electron mass splitting, we must choose $\cos^4 \theta_e \sim 1$. Therefore, $B=1/2$ and the ATOMKI's constraint in Eq.(\ref{protophobia}) leads to the following form giving the neutron mixing angle as follows 
\begin{equation}
 \cos^4 \theta_n = 2.7 \times 10^{-4}, ~~ \theta_n = - 1.442.
\end{equation}
The constraint from Fornal-Grinstein's resolution in Eq.(\ref{99}) now becomes
\begin{equation}
\sin^3 2 \theta_n \lambda^{10} = -0.261 \times 10^{24},
\end{equation}
which determines the geometric parameter $\lambda^2$ as follows
\begin{equation}
\lambda^2 = 4.8 \times 10^4.
\end{equation} 

Using this value of the geometric parameter $\lambda^2$ in Eq.(\ref{msplit}) we can calculate the neutron mass splitting  
\begin{equation}
m_{n}- m_{n_X} = 2.79~ keV. 
\end{equation}

All the decay channels in Eqs.(\ref{nnxdecay}) are energetically allowed if the mass of KK-neutrino has the upper bound
\begin{equation}
m_{\nu_X} < 1.38~keV.
\end{equation}

Thus, the mixing angle of neutrino has the following lower bound 
\begin{equation}
\sin 2 \theta_\nu =  \frac{2m}{\delta m_\nu \lambda^2} > 0.5133,~~ \theta_\nu > 0.27. 
\end{equation}

On the other hand, since $\sin 2 \theta_\nu < 1$, now we have a very narrow interval for $m_{\nu_X}$ as follows
\begin{equation}
1.38~keV > m_{\nu_X} \sim \delta m_\nu > \frac{2m}{\lambda^2} > 0.71~keV.
\end{equation}

The mass of the scalar boson $H$ represented by the field $h(x)$ in this case is predicted as
\begin{equation}
m_h = \frac{\sqrt{2} m}{\lambda^2} = 0.5~keV.
\end{equation}

{\it Large neutrino mass splitting}-- Now we can consider another extreme case when the mass of KK-neutrino is relatively large. For simplicity, we can choose $m_{\nu_X} > 17~MeV$ to have the ATOMKI's and Fornal-Grinstein's resolution constraints as follows
\begin{subequations} \label{largenu}
	\begin{alignat}{3}
	&&\frac{\cos^4 \theta_n}{1 + R} = 1.35 \times 10^{-4},~~ R= 1.002 \frac{\cos^4 \theta_\nu}{\cos^4 \theta_e} \label{ATOMKIlargenu} \\
	&&\sin^3 2 \theta_n \lambda^{10} = -0.261 \times 10^{24} \cos^4 \theta_\nu, \label{Fo-Grlargenu}
	\end{alignat}
\end{subequations}
where the physical meaning of $R$ is the ratio between two decays $X17 \rightarrow e^+ + e^-$ and $X17 \rightarrow \nu + \bar \nu$. In this scenario, the only alternative channel of neutron decay is $n \rightarrow n_X + \nu + \bar \nu$ as suggested by Ivanov et al \cite{Ivanov2019}. 

Since $ 0 > \sin 2 \theta_n > -1 $, Eqs.(\ref{largenu}) imply that $\cos^4 \theta_n$ and $R$ can vary in the following intervals
\begin{equation}
1 > \cos^4 \theta_n > \frac{1}{4},~~  7393 > R > 1852 .
\end{equation}
The physical meaning of these relations is that the coupling constant of $X17$ to electron in this case can be small, but its magnitude is $43-86$ times smaller than the one to neutrino to keep the ATOMKI's constraint.  

Eqs.(\ref{largenu}) must also be consistent with the condition $m_{\nu_X} > 17~MeV$ in this scenario. Therefore, we can use $R$ and $\delta m_n$ as variables to carry out the numerical estimation of the physical quantities by the following procedure: 
With given values of $R$, the neutron mixing angle $\theta_n$ and $\sin 2 \theta_n$ can be calculated based on Eq.(\ref{ATOMKIlargenu}). Then with a given $\delta m_n$, one can also determine the geometric parameter $\lambda^2$ based on the mass splitting formula (\ref{msplit}). Finally, based on Eq.(\ref{Fo-Grlargenu}, one can determine $\theta_\nu$ and $\delta m_\nu$. The condition $\delta m_{\nu_X} > 17~MeV$ will limit the possible values of $\delta m_n$. Finally, with the value of $\cos^4 \theta_\nu$ at hand at each value of $R$, the value of $\cos^4 \theta_e$ is also given. Hence one can calculate the electron mass splitting $\delta m_e$. Some typical numerical results are given in the following table
\begin{widetext}
\begin{center}
\begin{table}[h]
	\centering
	\begin{tabular}{|l|c|c|c|c|c|c|c|}
		\hline
		\hline
		 & $\theta_n$ & $m_{n_X}$ (MeV)& $\lambda^2$ & $m_{\nu_X}$ (MeV) & $m_{e_X}$(MeV) & $\cos^2 \theta_e $ \\
		\hline
	   Upper value  & $ -\pi/2$  & 938.468  & 30 & 5658 &37000 & $0.23 \times 10^{-9}$  \\
	   Lower value &$-\pi/2$ & 939.485  & 399 & 17.1 &112 & $1.45 \times 10^{-7}$ \\
		\hline 
		Upper value & $-\pi/12$  & 938.468 & 62 & 1139 &10208 & $0.728 \times 10^{-9}$  \\
		Lower value & $-\pi/12$  & 939.4  & 399  &17.1  &153 &$0.774 \times 10^{-7}$  \\
		\hline
		Upper bound & $ -\pi/81$  & 938.468 & 1000 & 17.2 & 159 & $0.72\times 10^{-7}$   \\
		Lower bound & $- \pi/81$  & 938.47 & 1010  & 17.1 & 159& $0.723 \times 10^{-7}$ \\
		\hline
		\hline
	\end{tabular}
	\caption{KK-particle masses and X17 coupling to electron in the large neutrino mass splitting scenario}
	\label{tab:template}
\end{table}
\end{center}
\end{widetext}

In this case, $X17$ is electrophobic, since its coupling to electron is in the order of magnitude of $10^{-9}- 10^{-7}$ times the electromagnetic one. The neutron mixing angle $\theta_b$ reaches its maximum value around $-\pi/81$. Here the masses of KK particles, the geometric parameter $\lambda$ and the electron mixing angle are determined. At smaller values of $\theta_n$, the model predicts KK-particle masses in the currently accessible energy ranges. The mass of the scalar boson $H$ is in the range of $60.25-793.23~keV$. These implications can also be verified by experiments at currently accessible energy scale.
  
{\it Summary and discussions} --
In this article, we probe an space-time structure, which is an extension of the usual ${\cal M}^4$ with a DED having two points at a distance $11.6~fm$ from each other. The vector boson $X17$ emerges naturally as the Kaluza-Klein partner of photon, receiving a mass of $m=17~MeV$ due to an abelian Higgs mechanism from a scalar $h(x)$, which is also a KK-partner of photon. In this space-time, the fermions are mixed to form the mass eigenstates with mass splitting depending on the mixing angles. This special but natural property makes this model predictive because it relate the mass splitting with the coupling of $X17$ to the given fermion type. At  the energy scale of nuclear physics, we consider a particle model of nucleon, electron and neutrino in this extended space-time. The ATOMKI experimental results and Fornal-Grinstein resolution to the neutron life time puzzle are consistent with this model. The model also has some interesting predictions in two different scenarios.

In the first scenario, the KK-neutrino mass is in range $ 1.38~keV > m_{\nu_X} > 0.71~keV$. The KK-neutron mass is $939.29~MeV$. Therefore, the new neutron decay channels into $n_X$, neutrino and its KK-sibling via a virtual $X17$ exchange can explain quantitatively the neutron life time puzzle. The coupling of $X17$ to electron is larger than $1.162 \times 10^{-2}$ times of the electromagnetic one. The masses of KK-proton and KK-neutron in this case are large, at least in the TeV range. In the second scenario, the coupling of $X17$ to electron is extremely small in the range of $10^{-9}-10^{-7}$ times the electromagnetic constant. The KK-electron mass in this case is limited in the range of $100~MeV-37~ GeV$. The KK-neutrino mass is in the range $17~MeV-5.7~GeV$. The KK-neutron mass is in the range $ 938.47-939.4~MeV$. All these predictions are verifiable at low energy.It is also remarkable that the model is consistent with the experimental results only if a non-trivial metric parameter $\lambda^2 > 30 $ is introduced. 

Since nucleons are composite particle, the above model can be built at the quark level with more accurate quantitative calculations. In the space-time with DED of size $11.6~fm$, the nucleon with radius less than $1~fm$ can be considered as a point-like particles in a good approximation. The predictions might be quantitatively modified slightly by using more precise value of $X17$ mass but most qualitative conclusions do not change. In our analysis, to be consistent with the experimental observations by Sun et al \cite{Sun} we have made an assumption, which is stronger than what suggested by Fornal and Grinstein \cite{FoGr} to suppress completely the $e^+-e^-$ pair production in the neutron decay. It can be argued that the possibility of allowing this pair with kinetic energy under $100~keV$, which is beyond the experiment detection ability, is still not completely closed. If this scenario is reconsidered, then KK-electron mass can be pushed up to $500~TeV$ by the above analysis. The scalar KK-partner $H$ of photon has a mass of $0.5~keV$ in the first scenario and in the range of $60.25-793.23~keV$ in the second one. The existence of $X17$ and $H$ as a consequence of the space-time extended by a discrete extra dimension considered in this article can also influence the muon magnetic momentum, proton radius, nucleon-nucleon and electron-nucleon scatterings. Additionally the contribution of $H$ in various physical processes can also be studied. These issues are worth to be discussed more extensively and currently are under progress.

Last but not least, by these results we have demonstrated that the extra dimension is not the exclusive task of the $TeV$ and Planck scale physics. Its indication can be searched in nuclear physics as well as in condensed matter physics as suggested before \cite{JVWQHE}.\\

{\it Acknowledgement}--Thanks are due to Tran Minh Hieu, Nguyen Van Dat and Pham Tien Du for their participation in the research group. In particular, I am grateful to Tran Minh Hieu, for reading the manuscript carefully. I would like to express my sincere thanks to ATOMKI and A.Krasznahorkay for their hospitality in Debrecen and continuous communications during my work. 

The supports of ITI, VNU and Department of Physics, Hanoi University of Science, VNU are greatly appreciated. The research is funded by Vietnam National Foundation for Science and Technology Development(NAFOSTED) under grant No 103.01-2017.319.

\end{document}